\documentclass[aps,pre,twocolumn,groupedaddress]{revtex4-2}
\usepackage{xcolor}
\usepackage{graphicx}
\usepackage[ruled,vlined]{algorithm2e}
\usepackage{amsmath}

\begin{document}


\title{An Information-Theoretic Law Governing Human Multi-Task Navigation Decisions}


\author{Nicholas Sohre$^{1}$}
\email[]{sohre007@umn.edu}
\author{Alisdair O. G. Wallis$^{2}$}
\author{Stephen J. Guy$^{1}$}
\email[]{sjguy@umn.edu}
\affiliation{Dept. of Computer Science \& Engineering, University of Minnesota, Minneapolis, MN 55455, USA$^{1}$ \\ Tesco PLC, Tesco House, Shire Park, Kestrel Way, Welwyn Garden City, AL7 1GA, UK$^{2}$}


\date{\today}

\begin{abstract}
To better understand the process by which humans make navigation decisions when tasked with multiple stopovers, we analyze motion data captured from shoppers in a grocery store. We discover several trends in the data that are consistent with a noisy decision making process for the order of item retrieval, and decompose a shopping trip into a sequence of discrete choices about the next item to retrieve. Our analysis reveals that the likelihood of inverting any two items in the order is monotonically bound to the information-theoretic entropy of the pair-wise ordering task. Based on this analysis, we propose a noisy distance estimation model for predicting the order of item retrieval given a shopping list. We show that our model theoretically reproduces the entropy-governed trend seen in the data with high accuracy, and in practice matches the trends in the data when used to simulate the same shopping lists. Our approach has direct applications to improving simulations of human navigation in retail and other settings.

\end{abstract}


\maketitle


\section{Introduction\label{intro}}
Understanding human flow through indoor buildings is important for various layout design tasks such as evacuation planning, product placement, and security. Advancements in technologies such as computer vision and motion-tracking have enabled the large scale collection long-term motion data, enabling new analyses aimed at incorporating these high level decisions into human flow models. Here, we take a data-driven approach to analyzing the navigation decisions of shoppers in a grocery store. 

\textit{Path data}, or varying spatial configurations of individuals as a function of time, provides valuable insight into human navigational behavior in a variety contexts~\cite{hui2009path}. Various works have utilized path data both as a means to understand and simulate human behavior, using different assumptions and conceptualizations to achieve specific research goals. Some of these works involve learning to predict human behavior from path data, from inferring high level flow models for retail floor optimization~\cite{ying2019customer}, to frameworks based on random walks~\cite{gutierrez2016active}, to crowd simulations~\cite{karamouzas2014universal}. Other works focus on analysis of behaviors, such as comparing human route selection to the theoretical optimum~\cite{hui2009research}, or discovering behavioral patterns~\cite{larson2005exploratory,karamouzas2018crowd}. Often these analyses are coupled with various mobility models for predicting human movement, designed to operate on different scales, from global migratory patterns~\cite{riascos2012long} and city level traffic~\cite{camargo2019diagnosing,piovani2018measuring} to more local path planning~\cite{lima2016understanding,bailenson2000initial} and collision avoidance~\cite{karamouzas2014universal,van2011reciprocal,helbing1995social}. Less explored is the mid-level scale of multi-task planning, such as how fair-goers might visit multiple attractions, or how customers determine which item to pick up next from a shopping list.

In this work, we use path data to better understand the process by which human shoppers make decisions about navigation when shopping. We focus on characterizing the task of selecting a next item to retrieve, and perform multiple analyses that provide insights into high level features governing shopper decisions. While numerous external factors affect this process, such as building layout and other human factors (specific in-store attractions, unplanned purchases~\cite{massara2014impulse}, or tendency to follow the perimeter~\cite{farley1966stochastic}), we adopt a general formulation of a shopping trip as a series of decisions given a predefined list of items, and utilize a large dataset of  shopper trajectories to gain insights about how the navigation process may be described and modeled from the path data. 

The rest of this paper is structured as follows. In Section~\ref{sec:data}, we describe the dataset as well as preprocessing steps and give some formal definitions that support our analysis. In Section~\ref{sec:analysis} we model shopping trips as a series of discrete navigation decisions that produces an item sequence, identify several high level trends, and propose a measure of decision difficulty that governs the suboptimalities in the data when decomposed into pairwise decision tasks. In Section~\ref{sec:independence} we incorporate our findings into a general decision model for each step of a shopping trip. In Section~\ref{sec:simulation} and propose a stochastic decision model that is theoretically guaranteed to produce the same trends seen in the data, but in practice matches the trends with high accuracy.


\section{Dataset\label{sec:data}}

\begin{figure}
    \centering
    \includegraphics[width=0.7\columnwidth]{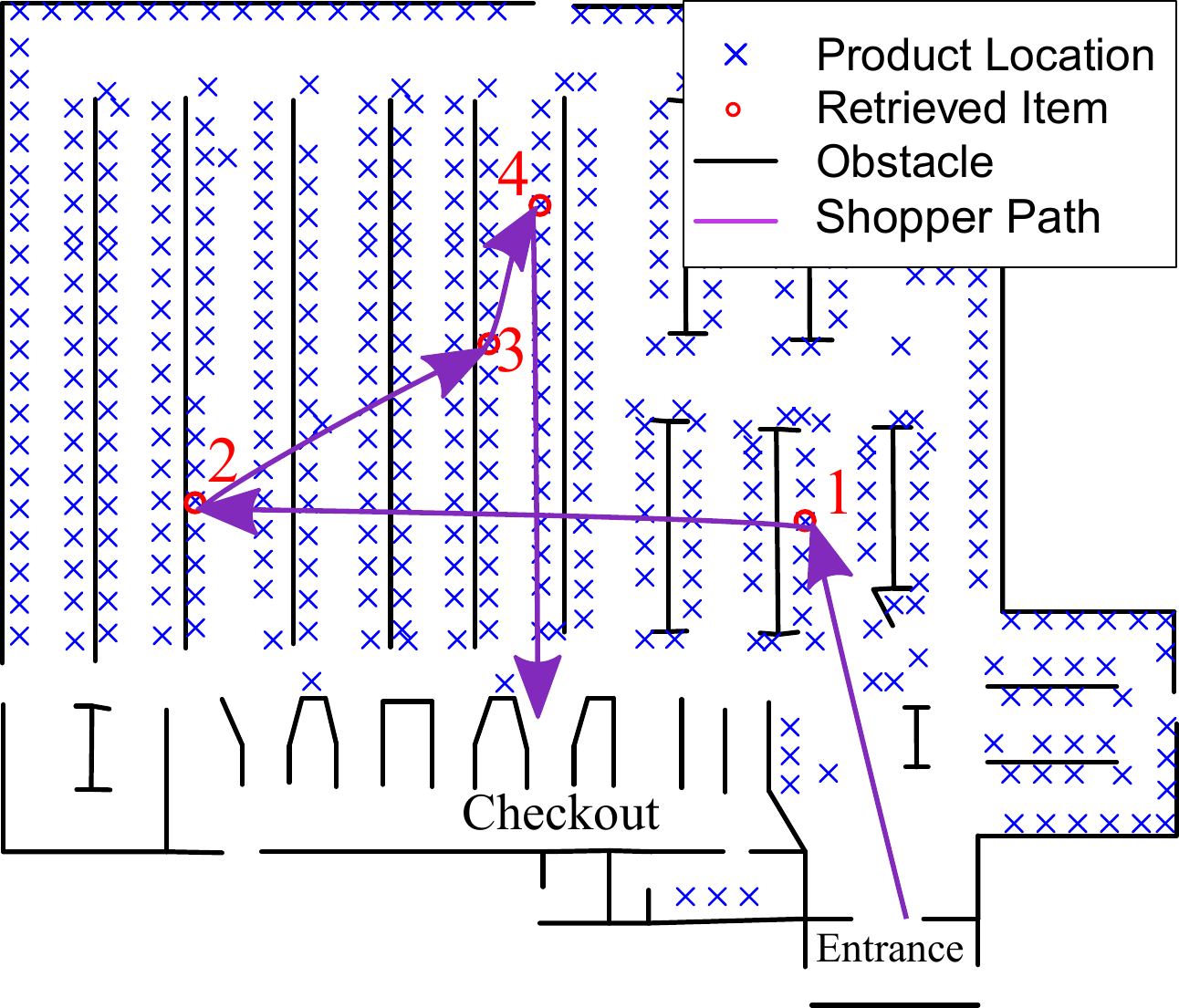}%
    \caption{(Color online) An example item sequence from the dataset, embedded in the abstracted store layout (black obstacles) and product shelf embedding (blue x's).\label{fig:entropy:example_match}}
\end{figure}

Here we use an anonymized dataset consisting of item sequences from shoppers in a retail store (\textit{paths}), corresponding to individual transactions from point-of-sale records of sets of items purchased together (\textit{baskets}). Each sequence reflects the order in which the items were retrieved. The items are embedded in a 2D representation of the store layout corresponding to product shelf locations. Additionally we use the set of 2D wall obstacles representing the sales floor layout to compute features such as walking distance between items. Figure ~\ref{fig:entropy:example_match} shows a contextualized example of a single shopping trip as it would appear in the data. The data contains over 13,000 such basket sequences spanning a period of two weeks.

We use these item orderings to derive a set of decision points for each shopping trip, where each decision point~$\mathbf{p}$ represents the set of remaining items the shopper eventually retrieves (but had not yet retrieved as of that point in the trip). 
%
The item sequences (and corresponding decision points) for each shopping trip represent paths over the fully connected item graph $\mathcal{G} = (\mathcal{V},\mathcal{E})$ where each vertex $v\in\mathcal{V}$ represents the spatial embedding of an item shelf in the store (we assign the shelf's position to be its corner), and the edges $e_{ij}\in\mathcal{E}$ represent the straight line connections between co-visible vertices $v_{i}$ and $v_{j}$. We augment this graph with an additional vertex $v_{start}$ for the entrance to the store to serve as the shopper location for the first decision point in a trip. We use this graph to compute the \textit{shortest path} walking distance between any two items for analysis.

Given a list of items left to collect $i \in \{1,2,...n\}$ we can decompose a decision point into the set of available (shortest walking path) travel distances it represents:
\begin{equation}
\mathbf{p} = \{d_{1},d_{2}...d_{n}\}
\label{eq:entropy:d}
\end{equation}
 
\noindent where each distance $d_i$ is the length of the shortest path over $\mathcal{G}$ from the item's shelf location to the shopper location (either $v_{start}$ or the vertex of the most recently chosen item's location).  We extract for analysis only those $\mathbf{p}$ containing more than one item, as having only a single item remaining does not present a choice to the shopper. The final set of decisions we use for analysis contains over 104,000 decision points. We use the distances in these decision points to predict shopper navigation choices.



\section{Data Analysis\label{sec:analysis}}

\subsection{Decisions}

 The navigation decisions in the data show that shoppers are generally efficient. Despite not necessarily having full knowledge of all item locations or store layout, about 79\% of the time, the next item a shopper picks up is the closest item to their current location $v_{j}$, having distance
$d^{*} = \text{min}_{i}(d_{i}\in\mathbf{p})$. While always picking the next closest item is not a globally optimal strategy (which involves solving a traveling salesperson problem, studied by~\cite{hui2009research} in the context of shopping), we refer to it as ``locally optimal" in the sense that it is the best path a person could take without knowing other future items they have not yet gathered (an approach suggested but not explored in~\cite{hui2009path}). This suggests the formulation of a shopping trip where each decision involves forming an estimate of which item left to collect is the closest, and navigating there. For the sake of analysis, we denote the next selected item for retrieval as $\hat{i}$, which is the index into a given $\mathbf{p}$ that reflects which item was chosen. Then $d_{\hat{i}}$ represents the distance to the chosen item. Similarly we define $i^{*}$ such that $d_{i^{*}} = d^{*}$ to be the index of the closest item for the decision point.

Due to the local nature of available information (and potentially limited familiarity shoppers may have with the store layout), cases where the closest item is not chosen ($\hat{i} \ne i^{*}$) naturally arise. The presence of these suboptimal local choices are consistent with other studies of human route selection and cognitive tasks where suboptimalities are found to be a natural part of these processes~\cite{lima2016understanding,hui2009research}. A suboptimal choice occurs whenever the chosen item's distance was larger than the optimal distance for that decision point. We call these choices \textit{inversions}, where the preference order of the chosen item with respect to the optimal one has been flipped. 

\begin{figure}
    \centering
    \includegraphics[width=0.95\columnwidth]{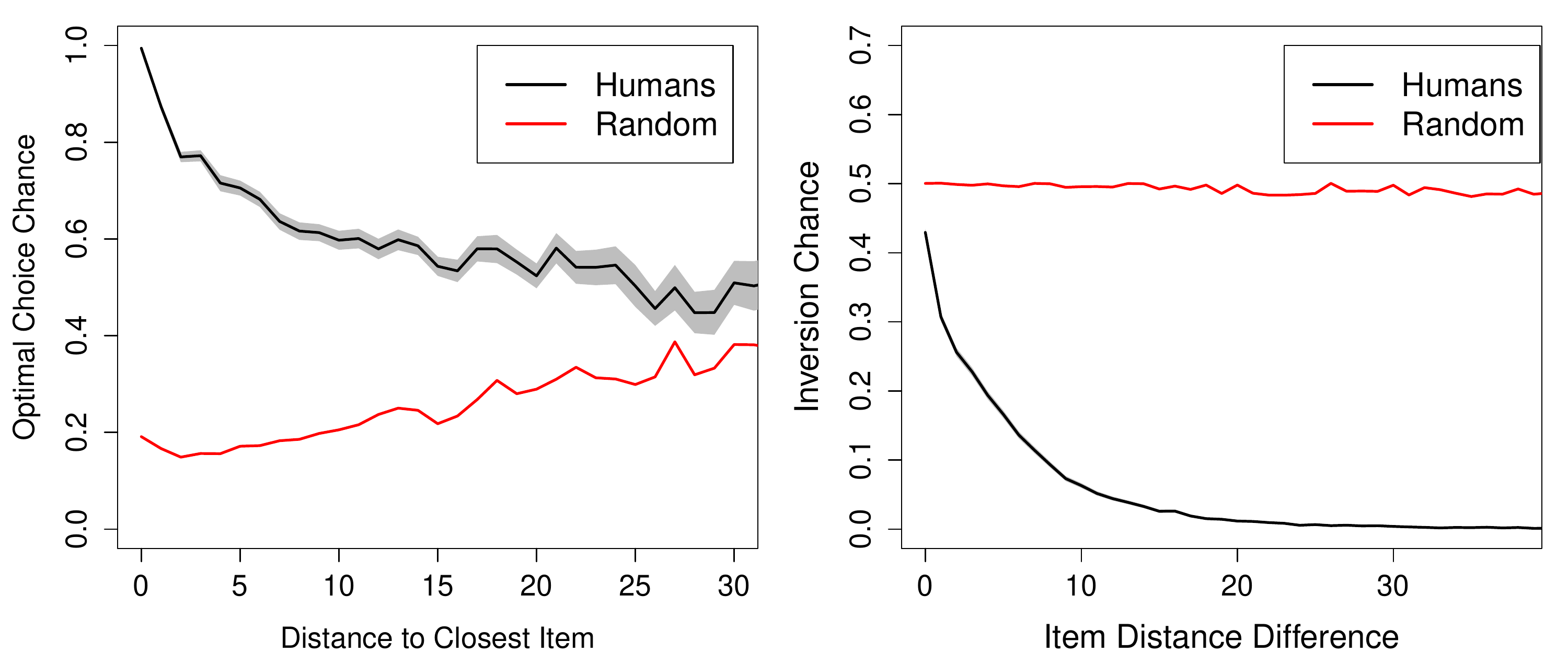}%
    \caption{(Color online) \textit{left}: The likelihood of choosing the locally optimal (closest) item decreases as a function of distance to the closest item for shopper paths (grey region is the 98\% confidence interval), and eventually converges to that of a random choice. \textit{right}: The likelihood of not choosing the closer of two items in a sub-task as a function of the difference in distances (to the shopper) between them. The random choice is shown for reference (always 0.5 for two items). \label{fig:inversion_trends}}
\end{figure}

An analysis of the inversions in the set of decision points reveals a strong trend involving $d^{*}$ and the likelihood of making a locally optimal choice (i.e., $\hat{d} = d^{*}$). Figure~\ref{fig:inversion_trends} (left) shows this trend, where the likelihood of choosing the optimal item at a decision point decreases as a function of increasing $d^{*}$. As this distance (and necessarily the distance of all other remaining items) increases, it becomes more difficult to consistently choose the closest item, converging toward the same likelihood as a random selection (shown in red).

\subsection{Sub-Tasks}
In addition to the set of per-decision inversions, we perform a decomposition of the shopping decisions to produce a larger set of inversions for analysis. To do this, we extract all pair-wise comparisons $(\hat{d},d_{i})$ from the data, each of which represents a possible inversion of the chosen item $\hat{d}_{i}$ with some alternative $d_{i}$. This yields a dataset of over $883,000$ such item pairs, which we refer to as \textit{sub-tasks}, in which a shopper estimates the closer of the two. 

The extracted pairs represent a subset of all the possible sub-tasks (and potential pair-wise inversions) that exist in the full decisions. Sub-task samples having $\hat{d} > d_{i}$ constitute pair-wise inversions, whose inversion amount can be measured as how much farther the shopper traveled then they would have by choosing item $i$. A sub-task can be alternatively decomposed as the pair $(F,C)$, where $F = \text{max}(\hat{d},d_{i})$ is the larger of the two, and $C = \text{min}(\hat{d},d_{i})$ is the smaller. This enables an inversion analysis based on both the closer item distance $C$ as well as the relative item distance $F-C$ (always a positive value). 

The right side of Figure~\ref{fig:inversion_trends} shows an analysis of the sub-tasks. Here, the chance of a pair-wise inversion falls off as the items grow farther apart in their relative distances to the shopper. This suggests that as the difference of relative distances grows, it becomes easier to distinguish which is closer.

\section{An Information-Theoretic Law for Inversion Likelihood}
\label{sec:inversion_law}
\subsection{Measuring Difficulty}
\label{sec:difficulty}
As is evident from the non-uniformity of both trends examined in Section~\ref{fig:inversion_trends}, some  decisions and sub-tasks are more likely to see inversions than others. Given the assumption that shoppers desire efficient paths, these trends serve as evidence that some scenarios present more difficult estimation tasks. Here we adopt a description of difficulty that follows from information theory, which is the \textit{Shannon entropy} of the sub-task. In the context of pair-wise decisions, the Shannon entropy is the minimum number of bits required to represent the item distances such that they can be reliably ordered. Formally, given the distance $F$ of the farther of the two items in a sub-task and the distance $C$ of the closer item, the entropy $H$ can be computed as

\begin{equation}
 \label{eq:entropy}
  \text{\textit{H}} = \log_2\left(\frac{F}{(F-C)}\right).
\end{equation}

We adopt the entropy of a sub-task as a measure of difficulty, and define the difficulty $D$ of a sub-task as:
\begin{equation}
 \label{eq:difficulty}
  \text{\textit{D}} = \log_2\left(\frac{F}{(F-C) + \epsilon}\right).
\end{equation}
\noindent where $\epsilon = 0.01$ places an upper bound on difficulty at $F=C$. We call $W = F-C$ the \textit{tolerance} of the task, as it is the maximum relative error in distance estimation that preserves their rank order. 
This difficulty measure is consistent with (and inspired by) those proposed in other psychophysical studies of human cognition~\cite{moyer1967time,fitts1964information}. 

\begin{figure}
    \centering
    \includegraphics[width=0.95\columnwidth]{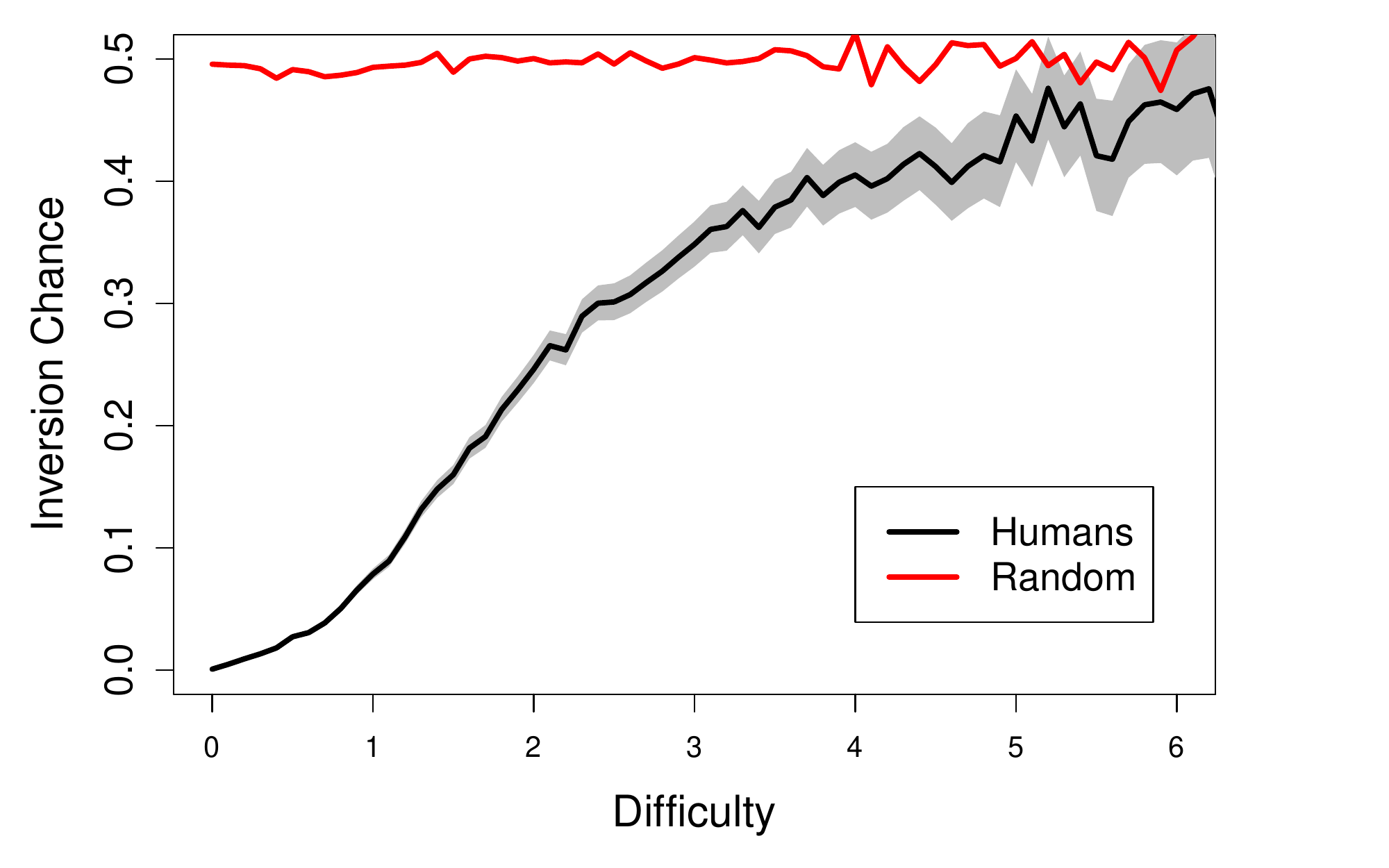}%
    \caption{(Color online) The chance of choosing the farther of two items in a sub-task as a function of difficulty (entropy) score for the shopper data (black w/ grey 98\% confidence interval) and random choice (red).\label{fig:D_data}}
\end{figure}

When inversion rate is graphed as a function of difficulty (Figure~\ref{fig:D_data}), we see a clear monotonic relationship between difficulty and inversion chance: $P(\hat{d} = F)$ (ie, choosing a farther than necessary item to go to next) increases as difficulty gets larger (with some noise affecting the trend at higher difficulties that occur less frequently in the data). Additionally, the inversion rates naturally converge with random selections at a maximum of 50\%. As the ability for shoppers to distinguish between closer and farther items saturates at around $D=5$, indicating that the information carrying capacity of the item selection process of shoppers for our data was around 5 bits.


This cognitive difficulty not only drives inversion rates, but also captures both empirical trends in Figure~\ref{fig:inversion_trends}. The closer together in distance from the shopper ($F-C$ is small), the larger $D$ becomes, consistent with Figure~\ref{fig:inversion_trends} (right) where there is greater confusion between item pairs having small distance differences. Additionally, the farther away the closer item is (large $F$), the greater $D$ becomes, matching the left side decision level trend showing lower chance of choosing optimally.

\subsection{Independent Perceptual Error Model\label{sec:independence}}
To extend the scope of the analysis in Section~\ref{sec:difficulty} to multi-item decisions, we note that the empirical trends for misordering items are consistent with a selection process that involves independent assessments of perceived item distances. We introduce $\Tilde{d}_{i}$ as a noisy, estimated distance to an item that incorporates uncertainty into a shopper's decision. The selection process can then be modeled as forming estimates for each item, then choosing the item $i$ that is estimated to be closest. In light of the analysis from Section~\ref{sec:analysis}, we note that the uncertainty of an item's estimated distance should grow with the true distance, and the ability to discriminate between them should diminish with close relative distances to the shopper. To meet these criteria, we design a generative model of noisy item distance estimation and model the noisy estimated distance, $\Tilde{d}_{i}$, as follows:
\begin{equation}
 \label{eq:gaussian_err}
  \Tilde{d}_{i} = d_{i} + \epsilon_{i},\; \epsilon_{i} \sim \mathcal{N}(0,\alpha d_{i})
\end{equation}
\noindent where the standard deviation $\alpha d_{i}$ is a linear function of the item's true distance, and each item's noise $w_{i}$ is sampled independently. For a pairwise sub-task, the chance of inversion can be computed directly from the gaussian noise model:

\begin{align}
P(\hat{d} > d_{i}) &= P(\hat{d} = F \;|\; F, C) \nonumber\\
 &= \frac{1}{2} 
 \left[ 
 1 + \text{erf}\left( 
 \frac{F - C}
 {\alpha \sqrt{2(C^2+F^2)}} 
 \right) 
 \right] 
\label{eq:gauss_inv}
\end{align}
\noindent where $F$ and $C$ represent the farther and closer distances of the sub-task $(\hat{d}, d_{i})$ respectively.

The supplementary materials provide a derivation of Equation \ref{eq:gauss_inv}. This analytical expression for the inversion chance both enables theoretical properties of the model to be guaranteed, and is efficient to compute, making it practical to directly fit $\alpha$ to the inversion chances in the data via numerical optimization techniques. Here we fit $\alpha$ to the trend shown in~\ref{fig:inversion_trends} (\textit{right}), as it has good data support over the entire domain (i.e., very tight confidence bounds across the x axis). Using BFGS gradient descent optimization, with a mean-squared error loss, yields $\alpha=0.30$ as a minimizing value. The optimization was performed using the \textit{optim} function of the \textit{stats} package for statistical computing language R~\cite{Rlanguage}.

\begin{figure}
    \centering
    \includegraphics[width=0.95\columnwidth]{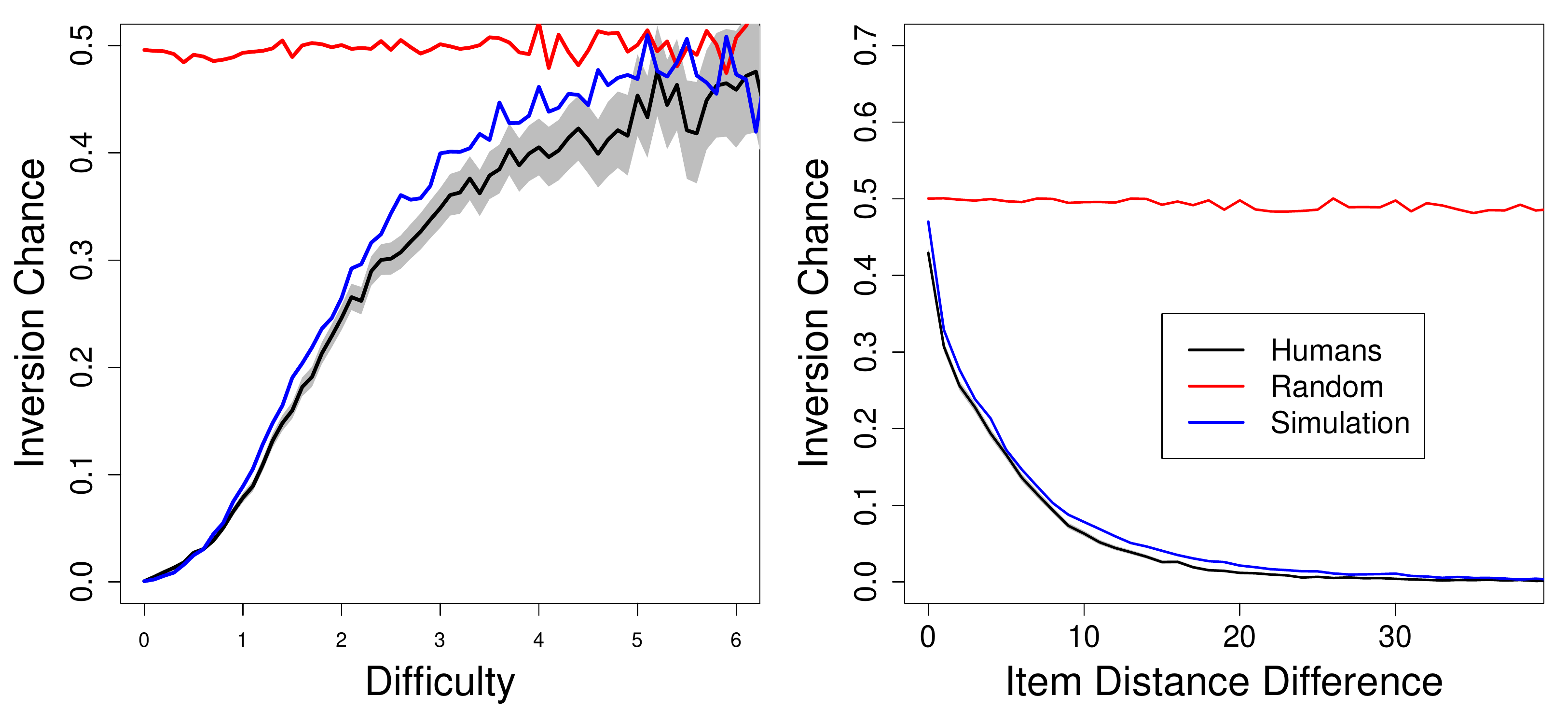}%
    \caption{(Color online) \textit{left}: the difficulty plot from Figure~\ref{fig:D_data} with overlaid simulation inversion rate for the independent distance estimation model (solid blue line)
    \textit{right}: observed inversion rates from simulated paths overlaid on the trend from Figure~\ref{fig:inversion_trends} (right). \label{fig:sub_task_level}}
\end{figure}

We can use Equation~\ref{eq:gauss_inv} to drive an agent-based simulation of navigation based on the basket data (See Section~\ref{sec:simulation} for further discussion of simulation details). Despite this predictive model using only a single parameter, the resulting simulated paths closely match the human data on a variety of metrics across a large span of item distances and decision difficulties (See Figure~\ref{fig:sub_task_level}).

Additionally, this noisy distance estimation model has key theoretical properties that guarantee the simulations will follow the same general trends seen in Figures~\ref{fig:inversion_trends} and~\ref{fig:D_data} independent of the choice of $\alpha$. First, the distribution for choosing between $N$ items converges to uniform as the relative distances become closer in magnitude (this can be seem by taking the limit as $F$ approaches $C$ in from equation~\ref{eq:gauss_inv}). Second, the chance of chance of pairwise inversions monotonically approaches the asymptote of 0.5 both as relative item-agent distances decrease and as distance to the closest item increases. Finally, our model provably recovers the monotonic relationship between difficulty and chance of inversion  (see the supplementary material for a proof).


\begin{figure}
    \centering
    \includegraphics[width=0.95\columnwidth]{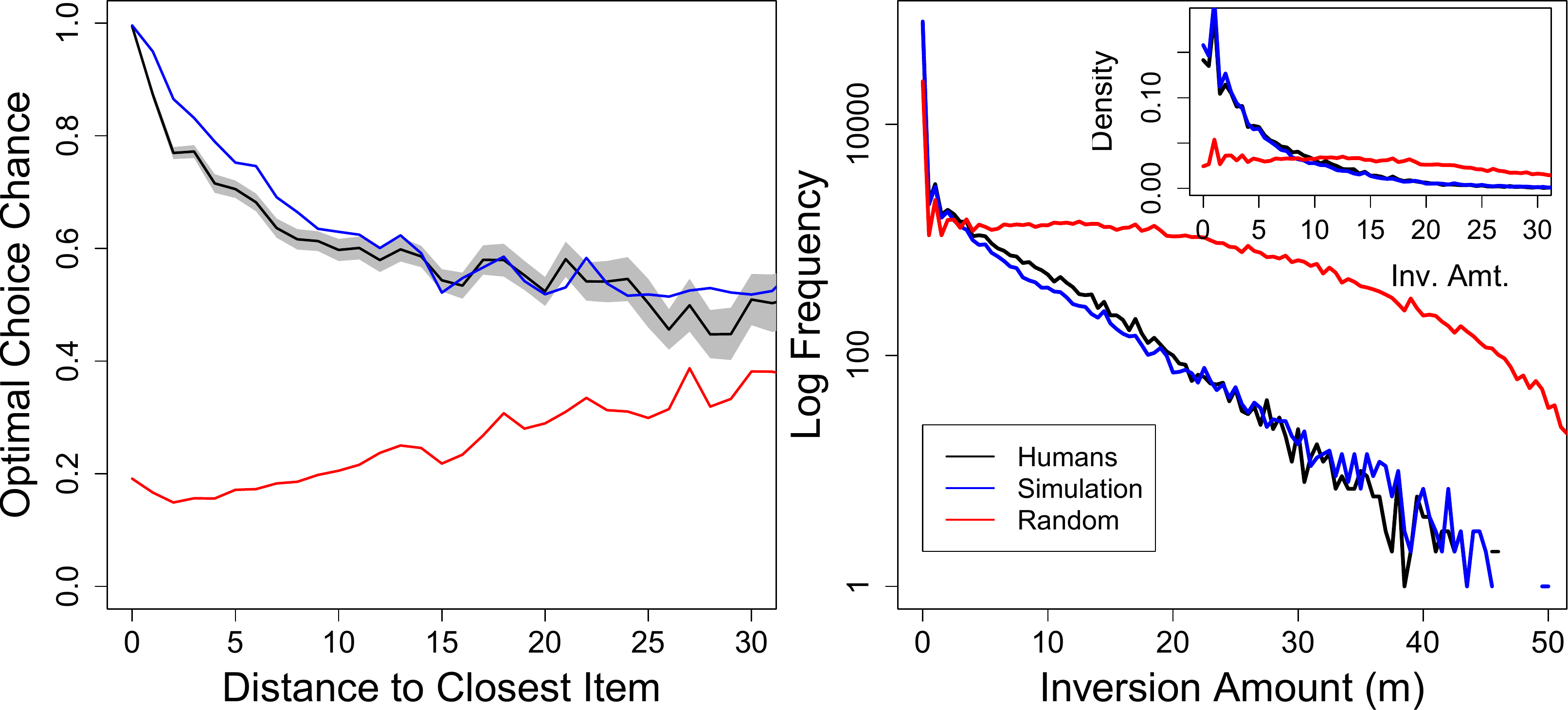}%
    \caption{(Color online)  \textit{left}: observed optimal choice rates from simulated paths overlaid on the same data from Figure~\ref{fig:inversion_trends} (left).
    \textit{right}: The log frequency of observed inversion sizes (ie, the extraneous distance travelled to the chosen item over the closest ) for simulated (blue) and shopper (black) item decisions. Inset is the density for inversion sizes cropped to show the knee of the curve.\label{fig:decision_level}}
\end{figure}

\section{Simulation Method\label{sec:simulation}}
The generative model from Section~\ref{sec:independence} is easily incorporated into a simulation for predicting a shopping trip given a basket of items to be collected, a spatial embedding of the items, and obstacles representing a store layout. We propose a stochastic agent-based simulation that does this with an execution strategy as follows. The agent begins at $v_{start}$ and has available a list of all $(d_{i},v_{i})$ tuples corresponding to where each item in the list can be picked up in the store. Noise is independently sampled according to equation~\ref{eq:gaussian_err} to produce a $\Tilde{d}_{i}$ for each candidate $(v_{i})$, and the agent selects the one having the smallest estimated distance as the next navigation target. The agent then navigates to the chosen item location using a shortest path planner over a visibility graph of the item locations. This strategy is repeated until all items have been collected (see the supplementary material for details).


To validate the simulation's accuracy at the decision level, we compare both the likelihood of inversions in the simulations to the shopper data, as well as the inversion magnitudes.
Figure~\ref{fig:decision_level} (left) shows the simulation results for the decision-level data trend from Section~\ref{sec:analysis}, where the likelihood of choosing the optimal item (the complement of inversion chance) matches well the shopper data compared to a random choice. On the right is a comparison of the log frequencies of \textit{inversion amounts} for the decisions, defined as $\hat{d} - d^{*}$. An inversion amount describes how much farther the chosen item was to the shopper than the closest item for a decision. Here, the simulation method continues to show good alignment with the shopper data, matching the actual inversion amounts with high accuracy.

The parameter $\alpha$ can be fit to data or used to tune behavior (for example, $\alpha \rightarrow 0$ will approach purely locally optimal decisions, which could be used to emulate shopper familiarity with the store layout). Additionally, the simulation technique is agnostic of store layout, and may be directly extended (or re-fit given new data) on any new store layout while retaining all the same properties that well describe the observed human behaviors.

\section{Conclusions\label{conclusions}}
In this work, we have performed a novel analysis of shopper path data to gain insight about multi-task human navigation decisions. When viewed as a sequence of decisions among remaining items, the data shows that shoppers very typically choose the next closest item to retrieve. The chance of item selection inversions in the data (that is, going to an item that was not the closest) follows our proposed sub-task difficulty measure monotonically. We observe that an independent error estimation model with a linear relationship between a true item's distance and uncertainty well captures these trends. Based on these findings we propose an agent-based method for simulating the order of item retrieval given a basket and a store layout. The simulated data recovers the relationship between the chance of inversions and sub-task difficulty, and in practice well matches the shopper data along several comparison metrics.
\vspace{1em}

\begin{acknowledgments}
This work was supported in part by the National Science Foundation under grants IIS-1748541 and CHS-1526693. 
\end{acknowledgments}

\bibliography{apssamp}

\providecommand{\noopsort}[1]{}\providecommand{\singleletter}[1]{#1}%
\begin{thebibliography}{19}%
\makeatletter
\providecommand \@ifxundefined [1]{%
 \@ifx{#1\undefined}
}%
\providecommand \@ifnum [1]{%
 \ifnum #1\expandafter \@firstoftwo
 \else \expandafter \@secondoftwo
 \fi
}%
\providecommand \@ifx [1]{%
 \ifx #1\expandafter \@firstoftwo
 \else \expandafter \@secondoftwo
 \fi
}%
\providecommand \natexlab [1]{#1}%
\providecommand \enquote  [1]{``#1''}%
\providecommand \bibnamefont  [1]{#1}%
\providecommand \bibfnamefont [1]{#1}%
\providecommand \citenamefont [1]{#1}%
\providecommand \href@noop [0]{\@secondoftwo}%
\providecommand \href [0]{\begingroup \@sanitize@url \@href}%
\providecommand \@href[1]{\@@startlink{#1}\@@href}%
\providecommand \@@href[1]{\endgroup#1\@@endlink}%
\providecommand \@sanitize@url [0]{\catcode `\\12\catcode `\$12\catcode
  `\&12\catcode `\#12\catcode `\^12\catcode `\_12\catcode `\%12\relax}%
\providecommand \@@startlink[1]{}%
\providecommand \@@endlink[0]{}%
\providecommand \url  [0]{\begingroup\@sanitize@url \@url }%
\providecommand \@url [1]{\endgroup\@href {#1}{\urlprefix }}%
\providecommand \urlprefix  [0]{URL }%
\providecommand \Eprint [0]{\href }%
\providecommand \doibase [0]{https://doi.org/}%
\providecommand \selectlanguage [0]{\@gobble}%
\providecommand \bibinfo  [0]{\@secondoftwo}%
\providecommand \bibfield  [0]{\@secondoftwo}%
\providecommand \translation [1]{[#1]}%
\providecommand \BibitemOpen [0]{}%
\providecommand \bibitemStop [0]{}%
\providecommand \bibitemNoStop [0]{.\EOS\space}%
\providecommand \EOS [0]{\spacefactor3000\relax}%
\providecommand \BibitemShut  [1]{\csname bibitem#1\endcsname}%
\let\auto@bib@innerbib\@empty
\bibitem [{\citenamefont {Hui}\ \emph {et~al.}(2009{\natexlab{a}})\citenamefont
  {Hui}, \citenamefont {Fader},\ and\ \citenamefont {Bradlow}}]{hui2009path}%
  \BibitemOpen
  \bibfield  {author} {\bibinfo {author} {\bibfnamefont {S.~K.}\ \bibnamefont
  {Hui}}, \bibinfo {author} {\bibfnamefont {P.~S.}\ \bibnamefont {Fader}},\
  and\ \bibinfo {author} {\bibfnamefont {E.~T.}\ \bibnamefont {Bradlow}},\
  }\bibfield  {title} {\bibinfo {title} {Path data in marketing: An integrative
  framework and prospectus for model building},\ }\href@noop {} {\bibfield
  {journal} {\bibinfo  {journal} {Marketing Science}\ }\textbf {\bibinfo
  {volume} {28}},\ \bibinfo {pages} {320} (\bibinfo {year}
  {2009}{\natexlab{a}})}\BibitemShut {NoStop}%
\bibitem [{\citenamefont {Ying}\ \emph {et~al.}(2019)\citenamefont {Ying},
  \citenamefont {Wallis}, \citenamefont {Beguerisse-D{\'\i}az}, \citenamefont
  {Porter},\ and\ \citenamefont {Howison}}]{ying2019customer}%
  \BibitemOpen
  \bibfield  {author} {\bibinfo {author} {\bibfnamefont {F.}~\bibnamefont
  {Ying}}, \bibinfo {author} {\bibfnamefont {A.~O.}\ \bibnamefont {Wallis}},
  \bibinfo {author} {\bibfnamefont {M.}~\bibnamefont {Beguerisse-D{\'\i}az}},
  \bibinfo {author} {\bibfnamefont {M.~A.}\ \bibnamefont {Porter}},\ and\
  \bibinfo {author} {\bibfnamefont {S.~D.}\ \bibnamefont {Howison}},\
  }\bibfield  {title} {\bibinfo {title} {Customer mobility and congestion in
  supermarkets},\ }\href@noop {} {\bibfield  {journal} {\bibinfo  {journal}
  {Physical Review E}\ }\textbf {\bibinfo {volume} {100}},\ \bibinfo {pages}
  {062304} (\bibinfo {year} {2019})}\BibitemShut {NoStop}%
\bibitem [{\citenamefont {Guti{\'e}rrez-Roig}\ \emph
  {et~al.}(2016)\citenamefont {Guti{\'e}rrez-Roig}, \citenamefont {Sagarra},
  \citenamefont {Oltra}, \citenamefont {Palmer}, \citenamefont {Bartumeus},
  \citenamefont {Diaz-Guilera},\ and\ \citenamefont
  {Perell{\'o}}}]{gutierrez2016active}%
  \BibitemOpen
  \bibfield  {author} {\bibinfo {author} {\bibfnamefont {M.}~\bibnamefont
  {Guti{\'e}rrez-Roig}}, \bibinfo {author} {\bibfnamefont {O.}~\bibnamefont
  {Sagarra}}, \bibinfo {author} {\bibfnamefont {A.}~\bibnamefont {Oltra}},
  \bibinfo {author} {\bibfnamefont {J.~R.}\ \bibnamefont {Palmer}}, \bibinfo
  {author} {\bibfnamefont {F.}~\bibnamefont {Bartumeus}}, \bibinfo {author}
  {\bibfnamefont {A.}~\bibnamefont {Diaz-Guilera}},\ and\ \bibinfo {author}
  {\bibfnamefont {J.}~\bibnamefont {Perell{\'o}}},\ }\bibfield  {title}
  {\bibinfo {title} {Active and reactive behaviour in human mobility: the
  influence of attraction points on pedestrians},\ }\href@noop {} {\bibfield
  {journal} {\bibinfo  {journal} {Royal Society open science}\ }\textbf
  {\bibinfo {volume} {3}},\ \bibinfo {pages} {160177} (\bibinfo {year}
  {2016})}\BibitemShut {NoStop}%
\bibitem [{\citenamefont {Karamouzas}\ \emph {et~al.}(2014)\citenamefont
  {Karamouzas}, \citenamefont {Skinner},\ and\ \citenamefont
  {Guy}}]{karamouzas2014universal}%
  \BibitemOpen
  \bibfield  {author} {\bibinfo {author} {\bibfnamefont {I.}~\bibnamefont
  {Karamouzas}}, \bibinfo {author} {\bibfnamefont {B.}~\bibnamefont
  {Skinner}},\ and\ \bibinfo {author} {\bibfnamefont {S.~J.}\ \bibnamefont
  {Guy}},\ }\bibfield  {title} {\bibinfo {title} {Universal power law governing
  pedestrian interactions},\ }\href@noop {} {\bibfield  {journal} {\bibinfo
  {journal} {Physical review letters}\ }\textbf {\bibinfo {volume} {113}},\
  \bibinfo {pages} {238701} (\bibinfo {year} {2014})}\BibitemShut {NoStop}%
\bibitem [{\citenamefont {Hui}\ \emph {et~al.}(2009{\natexlab{b}})\citenamefont
  {Hui}, \citenamefont {Fader},\ and\ \citenamefont
  {Bradlow}}]{hui2009research}%
  \BibitemOpen
  \bibfield  {author} {\bibinfo {author} {\bibfnamefont {S.~K.}\ \bibnamefont
  {Hui}}, \bibinfo {author} {\bibfnamefont {P.~S.}\ \bibnamefont {Fader}},\
  and\ \bibinfo {author} {\bibfnamefont {E.~T.}\ \bibnamefont {Bradlow}},\
  }\bibfield  {title} {\bibinfo {title} {Research note—the traveling salesman
  goes shopping: The systematic deviations of grocery paths from tsp
  optimality},\ }\href@noop {} {\bibfield  {journal} {\bibinfo  {journal}
  {Marketing science}\ }\textbf {\bibinfo {volume} {28}},\ \bibinfo {pages}
  {566} (\bibinfo {year} {2009}{\natexlab{b}})}\BibitemShut {NoStop}%
\bibitem [{\citenamefont {Larson}\ \emph {et~al.}(2005)\citenamefont {Larson},
  \citenamefont {Bradlow},\ and\ \citenamefont
  {Fader}}]{larson2005exploratory}%
  \BibitemOpen
  \bibfield  {author} {\bibinfo {author} {\bibfnamefont {J.~S.}\ \bibnamefont
  {Larson}}, \bibinfo {author} {\bibfnamefont {E.~T.}\ \bibnamefont
  {Bradlow}},\ and\ \bibinfo {author} {\bibfnamefont {P.~S.}\ \bibnamefont
  {Fader}},\ }\bibfield  {title} {\bibinfo {title} {An exploratory look at
  supermarket shopping paths},\ }\href@noop {} {\bibfield  {journal} {\bibinfo
  {journal} {International Journal of research in Marketing}\ }\textbf
  {\bibinfo {volume} {22}},\ \bibinfo {pages} {395} (\bibinfo {year}
  {2005})}\BibitemShut {NoStop}%
\bibitem [{\citenamefont {Karamouzas}\ \emph {et~al.}(2018)\citenamefont
  {Karamouzas}, \citenamefont {Sohre}, \citenamefont {Hu},\ and\ \citenamefont
  {Guy}}]{karamouzas2018crowd}%
  \BibitemOpen
  \bibfield  {author} {\bibinfo {author} {\bibfnamefont {I.}~\bibnamefont
  {Karamouzas}}, \bibinfo {author} {\bibfnamefont {N.}~\bibnamefont {Sohre}},
  \bibinfo {author} {\bibfnamefont {R.}~\bibnamefont {Hu}},\ and\ \bibinfo
  {author} {\bibfnamefont {S.~J.}\ \bibnamefont {Guy}},\ }\bibfield  {title}
  {\bibinfo {title} {Crowd space: a predictive crowd analysis technique},\
  }\href@noop {} {\bibfield  {journal} {\bibinfo  {journal} {ACM Transactions
  on Graphics (TOG)}\ }\textbf {\bibinfo {volume} {37}},\ \bibinfo {pages} {1}
  (\bibinfo {year} {2018})}\BibitemShut {NoStop}%
\bibitem [{\citenamefont {Riascos}\ and\ \citenamefont
  {Mateos}(2012)}]{riascos2012long}%
  \BibitemOpen
  \bibfield  {author} {\bibinfo {author} {\bibfnamefont {A.~P.}\ \bibnamefont
  {Riascos}}\ and\ \bibinfo {author} {\bibfnamefont {J.~L.}\ \bibnamefont
  {Mateos}},\ }\bibfield  {title} {\bibinfo {title} {Long-range navigation on
  complex networks using l{\'e}vy random walks},\ }\href@noop {} {\bibfield
  {journal} {\bibinfo  {journal} {Physical Review E}\ }\textbf {\bibinfo
  {volume} {86}},\ \bibinfo {pages} {056110} (\bibinfo {year}
  {2012})}\BibitemShut {NoStop}%
\bibitem [{\citenamefont {Camargo}\ \emph {et~al.}(2019)\citenamefont
  {Camargo}, \citenamefont {Bright},\ and\ \citenamefont
  {Hale}}]{camargo2019diagnosing}%
  \BibitemOpen
  \bibfield  {author} {\bibinfo {author} {\bibfnamefont {C.~Q.}\ \bibnamefont
  {Camargo}}, \bibinfo {author} {\bibfnamefont {J.}~\bibnamefont {Bright}},\
  and\ \bibinfo {author} {\bibfnamefont {S.~A.}\ \bibnamefont {Hale}},\
  }\bibfield  {title} {\bibinfo {title} {Diagnosing the performance of human
  mobility models at small spatial scales using volunteered geographical
  information},\ }\href@noop {} {\bibfield  {journal} {\bibinfo  {journal}
  {Royal Society open science}\ }\textbf {\bibinfo {volume} {6}},\ \bibinfo
  {pages} {191034} (\bibinfo {year} {2019})}\BibitemShut {NoStop}%
\bibitem [{\citenamefont {Piovani}\ \emph {et~al.}(2018)\citenamefont
  {Piovani}, \citenamefont {Arcaute}, \citenamefont {Uchoa}, \citenamefont
  {Wilson},\ and\ \citenamefont {Batty}}]{piovani2018measuring}%
  \BibitemOpen
  \bibfield  {author} {\bibinfo {author} {\bibfnamefont {D.}~\bibnamefont
  {Piovani}}, \bibinfo {author} {\bibfnamefont {E.}~\bibnamefont {Arcaute}},
  \bibinfo {author} {\bibfnamefont {G.}~\bibnamefont {Uchoa}}, \bibinfo
  {author} {\bibfnamefont {A.}~\bibnamefont {Wilson}},\ and\ \bibinfo {author}
  {\bibfnamefont {M.}~\bibnamefont {Batty}},\ }\bibfield  {title} {\bibinfo
  {title} {Measuring accessibility using gravity and radiation models},\
  }\href@noop {} {\bibfield  {journal} {\bibinfo  {journal} {Royal Society open
  science}\ }\textbf {\bibinfo {volume} {5}},\ \bibinfo {pages} {171668}
  (\bibinfo {year} {2018})}\BibitemShut {NoStop}%
\bibitem [{\citenamefont {Lima}\ \emph {et~al.}(2016)\citenamefont {Lima},
  \citenamefont {Stanojevic}, \citenamefont {Papagiannaki}, \citenamefont
  {Rodriguez},\ and\ \citenamefont {Gonz{\'a}lez}}]{lima2016understanding}%
  \BibitemOpen
  \bibfield  {author} {\bibinfo {author} {\bibfnamefont {A.}~\bibnamefont
  {Lima}}, \bibinfo {author} {\bibfnamefont {R.}~\bibnamefont {Stanojevic}},
  \bibinfo {author} {\bibfnamefont {D.}~\bibnamefont {Papagiannaki}}, \bibinfo
  {author} {\bibfnamefont {P.}~\bibnamefont {Rodriguez}},\ and\ \bibinfo
  {author} {\bibfnamefont {M.~C.}\ \bibnamefont {Gonz{\'a}lez}},\ }\bibfield
  {title} {\bibinfo {title} {Understanding individual routing behaviour},\
  }\href@noop {} {\bibfield  {journal} {\bibinfo  {journal} {Journal of The
  Royal Society Interface}\ }\textbf {\bibinfo {volume} {13}},\ \bibinfo
  {pages} {20160021} (\bibinfo {year} {2016})}\BibitemShut {NoStop}%
\bibitem [{\citenamefont {Bailenson}\ \emph {et~al.}(2000)\citenamefont
  {Bailenson}, \citenamefont {Shum},\ and\ \citenamefont
  {Uttal}}]{bailenson2000initial}%
  \BibitemOpen
  \bibfield  {author} {\bibinfo {author} {\bibfnamefont {J.~N.}\ \bibnamefont
  {Bailenson}}, \bibinfo {author} {\bibfnamefont {M.~S.}\ \bibnamefont
  {Shum}},\ and\ \bibinfo {author} {\bibfnamefont {D.~H.}\ \bibnamefont
  {Uttal}},\ }\bibfield  {title} {\bibinfo {title} {The initial segment
  strategy: A heuristic for route selection},\ }\href@noop {} {\bibfield
  {journal} {\bibinfo  {journal} {Memory \& Cognition}\ }\textbf {\bibinfo
  {volume} {28}},\ \bibinfo {pages} {306} (\bibinfo {year} {2000})}\BibitemShut
  {NoStop}%
\bibitem [{\citenamefont {Van Den~Berg}\ \emph {et~al.}(2011)\citenamefont {Van
  Den~Berg}, \citenamefont {Guy}, \citenamefont {Lin},\ and\ \citenamefont
  {Manocha}}]{van2011reciprocal}%
  \BibitemOpen
  \bibfield  {author} {\bibinfo {author} {\bibfnamefont {J.}~\bibnamefont {Van
  Den~Berg}}, \bibinfo {author} {\bibfnamefont {S.~J.}\ \bibnamefont {Guy}},
  \bibinfo {author} {\bibfnamefont {M.}~\bibnamefont {Lin}},\ and\ \bibinfo
  {author} {\bibfnamefont {D.}~\bibnamefont {Manocha}},\ }\bibfield  {title}
  {\bibinfo {title} {Reciprocal n-body collision avoidance},\ }in\ \href@noop
  {} {\emph {\bibinfo {booktitle} {Robotics research}}}\ (\bibinfo  {publisher}
  {Springer},\ \bibinfo {year} {2011})\ pp.\ \bibinfo {pages}
  {3--19}\BibitemShut {NoStop}%
\bibitem [{\citenamefont {Helbing}\ and\ \citenamefont
  {Molnar}(1995)}]{helbing1995social}%
  \BibitemOpen
  \bibfield  {author} {\bibinfo {author} {\bibfnamefont {D.}~\bibnamefont
  {Helbing}}\ and\ \bibinfo {author} {\bibfnamefont {P.}~\bibnamefont
  {Molnar}},\ }\bibfield  {title} {\bibinfo {title} {Social force model for
  pedestrian dynamics},\ }\href@noop {} {\bibfield  {journal} {\bibinfo
  {journal} {Physical review E}\ }\textbf {\bibinfo {volume} {51}},\ \bibinfo
  {pages} {4282} (\bibinfo {year} {1995})}\BibitemShut {NoStop}%
\bibitem [{\citenamefont {Massara}\ \emph {et~al.}(2014)\citenamefont
  {Massara}, \citenamefont {Melara},\ and\ \citenamefont
  {Liu}}]{massara2014impulse}%
  \BibitemOpen
  \bibfield  {author} {\bibinfo {author} {\bibfnamefont {F.}~\bibnamefont
  {Massara}}, \bibinfo {author} {\bibfnamefont {R.~D.}\ \bibnamefont
  {Melara}},\ and\ \bibinfo {author} {\bibfnamefont {S.~S.}\ \bibnamefont
  {Liu}},\ }\bibfield  {title} {\bibinfo {title} {Impulse versus opportunistic
  purchasing during a grocery shopping experience},\ }\href@noop {} {\bibfield
  {journal} {\bibinfo  {journal} {Marketing letters}\ }\textbf {\bibinfo
  {volume} {25}},\ \bibinfo {pages} {361} (\bibinfo {year} {2014})}\BibitemShut
  {NoStop}%
\bibitem [{\citenamefont {Farley}\ and\ \citenamefont
  {Ring}(1966)}]{farley1966stochastic}%
  \BibitemOpen
  \bibfield  {author} {\bibinfo {author} {\bibfnamefont {J.~U.}\ \bibnamefont
  {Farley}}\ and\ \bibinfo {author} {\bibfnamefont {L.~W.}\ \bibnamefont
  {Ring}},\ }\bibfield  {title} {\bibinfo {title} {A stochastic model of
  supermarket traffic flow},\ }\href@noop {} {\bibfield  {journal} {\bibinfo
  {journal} {Operations Research}\ }\textbf {\bibinfo {volume} {14}},\ \bibinfo
  {pages} {555} (\bibinfo {year} {1966})}\BibitemShut {NoStop}%
\bibitem [{\citenamefont {Moyer}\ and\ \citenamefont
  {Landauer}(1967)}]{moyer1967time}%
  \BibitemOpen
  \bibfield  {author} {\bibinfo {author} {\bibfnamefont {R.~S.}\ \bibnamefont
  {Moyer}}\ and\ \bibinfo {author} {\bibfnamefont {T.~K.}\ \bibnamefont
  {Landauer}},\ }\bibfield  {title} {\bibinfo {title} {Time required for
  judgements of numerical inequality},\ }\href@noop {} {\bibfield  {journal}
  {\bibinfo  {journal} {Nature}\ }\textbf {\bibinfo {volume} {215}},\ \bibinfo
  {pages} {1519} (\bibinfo {year} {1967})}\BibitemShut {NoStop}%
\bibitem [{\citenamefont {Fitts}\ and\ \citenamefont
  {Peterson}(1964)}]{fitts1964information}%
  \BibitemOpen
  \bibfield  {author} {\bibinfo {author} {\bibfnamefont {P.~M.}\ \bibnamefont
  {Fitts}}\ and\ \bibinfo {author} {\bibfnamefont {J.~R.}\ \bibnamefont
  {Peterson}},\ }\bibfield  {title} {\bibinfo {title} {Information capacity of
  discrete motor responses.},\ }\href@noop {} {\bibfield  {journal} {\bibinfo
  {journal} {Journal of experimental psychology}\ }\textbf {\bibinfo {volume}
  {67}},\ \bibinfo {pages} {103} (\bibinfo {year} {1964})}\BibitemShut
  {NoStop}%
\bibitem [{\citenamefont {{R Core Team}}(2020)}]{Rlanguage}%
  \BibitemOpen
  \bibfield  {author} {\bibinfo {author} {\bibnamefont {{R Core Team}}},\
  }\href {https://www.R-project.org/} {\emph {\bibinfo {title} {R: A Language
  and Environment for Statistical Computing}}},\ \bibinfo {organization} {R
  Foundation for Statistical Computing},\ \bibinfo {address} {Vienna, Austria}
  (\bibinfo {year} {2020})\BibitemShut {NoStop}%
\end{thebibliography}%
\end{document}



\title{An Entropy Law Governing Retail Shopper Decisions: Supplemental Material}


\author{Nicholas Sohre$^{1}$}
\email[]{sohre007@umn.edu}
\author{Alisdair Wallis$^{2}$}
\author{Stephen J. Guy$^{1}$}
\email[]{sjguy@umn.edu}
\affiliation{University of Minnesota$^{1}$, Computer Science \& Engineering, Tesco PLC$^{2}$}

\maketitle
\appendix
\section{Proofs \label{app:proof}}
\paragraph{\textbf{Lemma:} Simulated Inversion rate of two items}
Given two items having true geodesic distances $b$ and $c$ from the agent, their estimated distances $\hat{b}$ and $\hat{c}$ under our simulation model are
\begin{equation}
\begin{split}
\hat{b} = b + w_{b}, \epsilon_{b} \sim \mathcal{N}(0,\alpha*b)\\
\hat{c} = c + w_{c}, \epsilon_{c} \sim \mathcal{N}(0,\alpha*c) 
\end{split}
\end{equation}
Which produce Gaussian random variables of the estimated distances $B \sim \mathcal{N}(b,\alpha*b) $ and $C \sim \mathcal{N}(c,\alpha*c)$.  Suppose $b < c$ (that is, item $b$ is closer to the agent than item $c$). Then, the probability of an inversion is the probability that the estimated distances swap in magnitude: $C < B \rightarrow C - B < 0$. Let $Y = C - B$ be a new Gaussian random variable, then 
\begin{equation}
\begin{split}
Y \sim \mathcal{N}(c-b, \sqrt{(\alpha*c)^2 + (\alpha*b)^2})\\
= \mathcal{N}(c-b, \alpha*\sqrt{c^2+b^2})
\end{split}
\end{equation}
and the likelihood of inversion is $P(Y < 0)$. For a given $b$, $c$, and $\alpha$, this quantity can be computed analytically from the CDF of $Y$ evaluated at $0$:
\begin{equation}
P(Y < 0) = \frac{1}{2} \left[ 1 + \text{erf}\left( \frac{0 - (c - b)}{\alpha*\sqrt{2(c^2+b^2)}} \right) \right] 
\label{eq:gauss_inv}
\end{equation}

\paragraph{\textbf{Theorem:} Simulated inversion chance increases with distance to the closer item}
First, we note that $1 + \text{erf}(x)$ is a positive, increasing function of $x$. Then, given equation~\ref{eq:gauss_inv}, it suffices to show that for any $b$ (the distance to the closer item), the input to \textit{erf} is increasing:
\begin{equation}
\forall b \left[ \frac{\delta}{\delta b}\, \frac{b - c }{\alpha*\sqrt{2(c^2+b^2)}} \geq 0\right] ,\;  0 < b \leq c
\label{eq:erf_inner}
\end{equation}

\textbf{Proof:} evaluating the partial derivative in equation~\ref{eq:erf_inner} with respect to $b$, we have

\begin{align}
& \frac{\delta}{\delta b}\, \frac{b - c }{\alpha*\sqrt{2(c^2+b^2)}}& \nonumber\\
& = \frac{1}{\alpha\sqrt{2}} \frac{\delta}{\delta b}\, \frac{b - c }{\sqrt{c^2+b^2}}& \nonumber\\
& = \frac{1}{\alpha\sqrt{2}} \frac{\sqrt{b^2+c^2} - \frac{(b-c)b}{\sqrt{b^2+c^2}}}{(b^2+c^2)}\;\; & \text{(quotient rule)} \nonumber\\
& = \frac{1}{\alpha\sqrt{2}}\frac{c(c+b)}{(b^2+c^2)^{\frac{3}{2}}}\;\; & \text{(simplify)}
\label{eq:inv_derivative}
\end{align}
Since equation~\ref{eq:inv_derivative} is positive whenever $b, c$ and $\alpha$ are positive, the derivative is positive with respect to $b$ and constrains equation~\ref{eq:gauss_inv} to be increasing with increasing $b$. 

\paragraph{\textbf{Theorem:} Simulated inversion chance decreases with increasing distance between items}
Another important property for maintaining the relationship between difficulty and inversion chance is that the inversion chance must decrease with increasing $W = c-b$.\\
\textbf{Proof:} We wish to show that the derivative with respect to $W$ is always negative:
\begin{equation}
\forall W = c-b, \left[ \frac{\delta}{\delta W}\, \frac{b - c }{\alpha*\sqrt{2(c^2+b^2)}} \leq 0\right] ,\;  0 < b \leq c
\label{eq:erf_inner_W}
\end{equation}

Noting that $(b^2 + c^2) = W^2 + 2cb = W^2 + 2b(b+W)$, we can substitute into equation~\ref{eq:erf_inner_W} and get
\begin{equation}
    \frac{\delta}{\delta W}\, \frac{-W}{\alpha*\sqrt{2(W^2 + 2b(b+W))}}
\label{eq:inv_derivative_subW}
\end{equation}

While we cannot write the derivative in terms of only $W$, we can treat $b$ as a positive constant and take the partial with respect to $W$. If the result is negative for any value of $b$, then the derivative with respect to $W$ is negative regardless of $b$ and the property is satisfied:
\begin{align}
& \frac{\delta}{\delta W}\, \left[-\frac{W}{\alpha*\sqrt{2(W^2 + 2b(b+W))}}\right]& \nonumber\\
& = \frac{1}{\alpha\sqrt{2}} \frac{\delta}{\delta W}\, \left[-\frac{W}{\sqrt{W^2 + 2b(b+W)}}\right]& \nonumber\\
& = \frac{1}{\alpha\sqrt{2}} \left[-\frac{\sqrt{W^2 + 2b(b+W)} - \frac{W(W+b)}{\sqrt{W^2 + 2b(b+W)}} }{W^2 + 2b(b+W)}\right]\;\; & \text{(quotient rule)} \nonumber\\
& = \frac{1}{\alpha\sqrt{2}} \left[ \frac{W(W+b)}{(W^2 + 2b(b+W))^\frac{3}{2}} - \frac{W^2+2b(b+W)}{(W^2 + 2b(b+W))^\frac{3}{2}}\right]\;\; & \text{(simplify)}\nonumber\\
& = \frac{1}{\alpha\sqrt{2}} \frac{-b(2b+W)}{(W^2 + 2b(b+W))^\frac{3}{2}}\;\; & \text{(simplify)}
\label{eq:inv_derivative_W}
\end{align}

Since $W$ and $b$ are both positive non-zero quantities, the resulting derivative is always negative as desired.  

\paragraph{\textbf{Theorem:} Simulation inversion chance increases monotonically with difficulty}
To show this property, it is sufficient to show that difficulty also increases monotonically with decreasing $W$ and increasing $b$, since both these two values fully specify both the inversion rate (given an $\alpha$) and the difficulty.\\
\textbf{Proof:} First, we note that $W$ and $b$ are sufficient to fully describe difficulty:
\begin{equation}
\text{\textit{difficulty}} = log_2\left(\frac{A}{W}\right) = log_2\left(\frac{ b+W}{W}\right)  
\end{equation}
Then we can construct the partial derivatives with respect to both $W$ and $b$ for difficulty and see that they are always positive and negative respectively:
\begin{equation}
\frac{\delta}{\delta W} \frac{A}{W} =\frac{\delta}{\delta W} \frac{b+W}{W} = \frac{-b}{W^2}
\end{equation}
\begin{equation}
\frac{\delta}{\delta b} \frac{A}{W} =\frac{\delta}{\delta b} \frac{b+W}{W} = \frac{b+W}{W^2}
\end{equation}
Thus, both inversion rate and difficulty monotonically increase with increasing $b$ and decrease with increasing $W$. Since $W$ and $b$ are both sufficient to fully specify both inversion rate and difficulty, we cannot make one smaller or larger (by adjusting $W$ or $b$) without having the same effect on the other. Therefore, both these quantities will have a monotonic relationship with each other.

\section{Simulation Details \label{app:sim}}
\begin{algorithm}[H]
\SetAlgoLined
\textbf{Input:} itemsToRetrieve\;
\textbf{Output:} itemOrder\;
 itemOrder = []\;
 alpha = 0.30\;
 \While{While itemsToRetrieve.length $<$ 1}{
  distances = []\;
  \For{i in 0 : itemsToRetrieve.length-1 }
  {
    trueDistance = getGeodesicDistance(itemsToRetrieve[i])\;
    noisyDistance = trueDistance + sampleNormal(0, alpha  * trueDistance)\;
    distances.push(noisyDistance)\;
  }
  itemsByEstimatedDist = sort(itemsToRetrieve, by = distances)\;
  itemOrder.push(itemsByEstimatedDist[0])\;
  itemsToRetrieve.remove(itemsByEstimatedDist[0])\;
 }
 itemOrder.push(itemsToRetrieve[0])\;
 return itemOrder;
 \caption{Basket Simulation \label{alg:simulation}}
\end{algorithm}